\documentclass[superscriptaddress,twocolumn,amsfonts,amsmath,amssymb,aps,prl,nofootinbib]{revtex4-1}
\usepackage[colorlinks=true,citecolor=blue,urlcolor=black]{hyperref}

\usepackage{graphicx}
\usepackage{dcolumn}
\usepackage{bm}
\usepackage{color}
\usepackage{subfigure}
\usepackage{amsthm}
\usepackage{amsmath}
\usepackage{amssymb}
\newtheorem{theorem}{Theorem}
\newtheorem{lemma}{Lemma}

\definecolor{Red}{rgb}{1,0,0}

\def\ket#1{| #1 \rangle}
\def\bra#1{\langle #1 |}

\def\Tr{\operatorname{Tr}}

\def\R{\mathbb{R}}

\begin{document}

\title{Exponential enhancement of quantum metrology using continuous variables}
\author{Li Sun}

\affiliation{Institute of Fundamental and Frontier Sciences, University of Electronic Science and Technology of China, Chengdu, Sichuan, 610051, China}

\author{Xi He}

\affiliation{Institute of Fundamental and Frontier Sciences, University of Electronic Science and Technology of China, Chengdu, Sichuan, 610051, China}

\author{Chenglong You}

\affiliation{Quantum Photonics Laboratory, Department of Physics \& Astronomy, Louisiana State University, Baton Rouge, LA 70803, USA}

\author{Chufan Lyu}

\affiliation{Institute of Fundamental and Frontier Sciences, University of Electronic Science and Technology of China, Chengdu, Sichuan, 610051, China}

\author{Bo Li}

\affiliation{The Key Laboratory of Mathematics Mechanization, Academy of Mathematics and Systems Science, Chinese Academy of Sciences, Beijing 100190, China}

\author{Seth Lloyd}
\email{slloyd@mit.edu}
\affiliation{Department of Mechanical Engineering, Massachusetts
Institute of Technology, Cambridge, Massachusetts 02139, USA}

\author{Xiaoting Wang}
\email{xiaoting@uestc.edu.cn}
\affiliation{Institute of Fundamental and Frontier Sciences, University of Electronic Science and Technology of China, Chengdu, Sichuan, 610051, China}

\date{\today}

\begin{abstract}

Coherence time is an important resource to generate enhancement in quantum metrology. In this work, based on continuous-variable models, we propose a new design of the signal-probe Hamiltonian which generates an exponential enhancement of measurement sensitivity. The key idea is to include into the system an ancilla that does not couple directly to the signal. An immediate benefit of such design is one can expand quantum Fisher information(QFI) into a power series in time, making it possible to achieve a higher-order time scaling in QFI. Specifically, one can design the interaction for a qubit-oscillator Ramsey interferometer to achieve a quartic time scaling, based on which, one can further design a chain of coupled harmonic oscillators to achieve an exponential time scaling in QFI. Our results show that linear scaling in both time and the number of coupling terms is sufficient to obtain exponential enhancement. Such exponential advantage is closely related to the characteristic commutation relations of quadratures.

\end{abstract}

\maketitle

\label{sec:Intro}
Quantum metrology aims to study the limitation of the measurement accuracy governed by quantum mechanics and to explore how to achieve better measurement sensitivity with quantum resources\cite{caves1981quantum, giovannetti2006quantum}. In recent years, quantum metrology has been long pursued due to its vital importance in applied physics, such as gravitational wave detection \cite{schnabel2010quantum, danilishin2012quantum, adhikari2014gravitational}, atomic clocks \cite{derevianko2011colloquium, schioppo2017ultrastable}, quantum imaging \cite{ kolobov1999spatial, dowling2015quantum, Omar2019}. A quantum metrology process includes three steps: preparing probes in a designed initial state; probes evolve under a parameter-dependent Hamiltonian $H_f$ for time $T$; the measurement of the final state.  A well-studied example of parameter estimation is to estimate a parameter $f$ introduced in a Hamiltonian with the form of $H_f=f H_{0}$, where $H_{0}$ is a known Hamiltonian. The sensitivity of estimating this type of parameter scales as $1/(T\Delta H_{0})$, where $\Delta H_{0}$ is the standard deviation of $H_{0}$. There are two ways to improve measurement accuracy: either by increasing the coherent interaction time $T$, or by maximizing the standard deviation of $H_{0}$ through preparing the probe in a special entangled state. Specifically, with quantum entanglement, one can maximize $\Delta H_{0}$ scaling as $\Delta H_{0} \propto N $, where $N$ is the number of probes \cite{giovannetti2006quantum}. Accordingly, the estimation sensitivity $ {1}/{(TN)}$ is considered as the Heisenberg limit, where $N$ can be perceived as the quantum parallel resource, and $T$ as the quantum serial resource \cite{huelga1997improvement, shaji2007qubit}.

Quantum metrology has been studied for a wide range of systems with quantum resource in parallel scheme\cite{giovannetti2006quantum, giovannetti2004quantum,  boixo2007generalized, boixo2008quantum, woolley2008nonlinear, anisimov2010quantum, giovannetti2011advances, thomas2011real, hall2012does, demkowicz2012elusive, PhysRevA.99.042122, arvidsson2020quantum, you2020multiphoton}. For instance, with $k$-body interactions between the probes, a sensitivity limit that scales as $1/N^{k}$ can be obtained \cite{boixo2007generalized}, while an exponential scaling can be achieved by introducing an exponentially large number of coupling terms~\cite{PhysRevLett.100.220501}. For quantum resource in serial scheme, in terms of the coherence time $T$, 
many questions remain open. It has been shown that the minimum sensitivity $\Delta f$ scales as $1/T$ with a time-independent Hamiltonian \cite{huelga1997improvement, yuan2015optimal, jones2020remote}, while $1/T^2$ scaling can be realized with a time-dependent Hamiltonian~\cite{pang2017optimal, naghiloo2017achieving}. Interesting open questions include: what is the ultimate limit of such enhancement, whether one can achieve sensitivity scaling $1/T^k$ for arbitrary $k$, or even the exponential scaling with the amount of other physical resources polynomial in $T$?

Here we show that exponential sensitivity can be efficiently achieved with the number of coupling terms scaling linear with time. Specifically, we study a special type of models with the Hamiltonian in the form $H_{f}=f H_{0}+H_{1}$, where $H_0$ is coupled with the signal $f$ and $H_{1}$ is an auxiliary Hamiltonian not coupled directly to $f$. Such model utilizes the non-commutativity of $H_0$ and $H_1$ to expand QFI into a power series in $T$, permitting us to obtain higher-order time scaling. In the model of a qubit-harmonic-oscillator(HO) Ramsey interferometer, sensitivity characterized by QFI with time scaling of $T^4$ can be achieved; in the second model with a chain of coupled harmonic resonators, the QFI can obtain an exponential improvement in the measurement accuracy. Remarkably, the second model only requires a polynomial (linear) scaling of coupling terms, which is crucial to justify the efficiency and the effectiveness of exponential enhancement. After all, it is of no surprise to realize exponential enhancement with an exponential amount of physical resource. 

\emph{Time resource for quantum enhancement.} --- In quantum metrology, one aims to estimate a parameter $f$ encoded in the quantum dynamics from $\nu$ repeated quantum measurements. The variance of the estimation or the measurement sensitivity $\Delta f$ is bounded by the quantum Cram\'er-Rao bound (QCRB): $\Delta f \ge 1/\sqrt{\nu F_{Q}}$, where the QFI ${F_Q} = \Tr(\rho _{f} L_f^2)$ corresponds to the minimum quantum measurement sensitivity\cite{braunstein1994statistical, braunstein1996generalized}. The parameter $f$ is first encoded into the Hamiltonian $H_f$, and then into the final state $\rho_f$ after an evolution time $T$ under $H_f$, with $\rho_f=U_f\rho_0U_f^\dag$ and $U_f=e^{-iH_fT}$. Here, $L_f$ is called the \emph{symmetric logarithmic derivative} (SLD), defined through the relation $\partial_ f {\rho_f} = \frac{1}{2}(\rho_{f} L_f+L_f \rho_{f})$. Alternatively, QFI can be reformulated in terms of $U_f$ and $\rho_0$. Specifically, for pure initial state $\rho_0=\ket{\psi_0}\bra{\psi_0}$, we have $F_Q = 4\Delta ^{2}h\equiv 4\bra{\psi_0 }h^2 \ket{\psi_0 }-|\bra{\psi_0}h \ket{\psi_0}|^2$, where $h\equiv i(\partial _{f}U_f^{\dag})U_f =-\int _{0}^{T} e^{i H_{f}t}{\partial _{f} H_{f}}e^{-i H_{f}t}dt$~\cite{holevo2011probabilistic, liu2015quantum}. It turns out that such reformulation of FQI is very useful to understand the scheme of quantum enhancement. For example, in the standard setting of quantum metrology where $H_f=f H_{0}$, the above formula gives $F_{Q}=4T^{2} \Delta^{2} H_{0}$. Hence, one way to improve QFI is to increase the coherent interaction time $T$. However, due to the decoherence and dephasing effects, $T$ cannot be extended arbitrarily long to improve QFI. In comparison, achieving higher-order terms of $T$ in QFI is a more efficient and practical method to improve the estimation precision. 

In order to find the relationship between $F_Q$ and $T$, we consider a general form of $H_f$ on $f$ and rewrite $h$ into the following polynomial expansion in $T$: 
\begin{align}
 h=-\sum _{j=0}^{\infty}\frac{i^j T^{j+1}}{(j+1)!}C_j,
\end{align}
where $C_j\equiv [H_f, C_{j-1}]$ is the $j$th-order commutator of $H_f$ and $\partial _f H_f$ with $C_0=\partial _{f} H_{f}$  \cite{kumar1965expanding, wilcox1967exponential, PhysRevA.90.022117, liu2015quantum}.
Thus, we can obtain the expression of QFI:
\begin{align}\label{Eq3}
	 F_{Q}= 4\Delta ^{2}\Big(\sum _{j=0}^{\infty}\frac{i^j T^{j+1}}{(j+1)!}C_j\Big). 
\end{align}
Notice that the series in $h$ could be of finite length if $C_j=0$ for some $j$. If $C_j$ are local operators on different subsystems, then for a separable initial state $\ket\psi$, we have the covariance ${\rm Cov } (C_k,C_j)\equiv \bra{\psi}C_kC_j\ket{\psi}-\bra{\psi}C_k\ket{\psi}\bra{\psi}C_j\ket{\psi}=0$, and the QFI can be further simplified into
\begin{align}\label{Eq31}
F_Q=4\sum _{j=0}^{\infty}\frac{ (-1)^jT^{2j+2}}{[(j+1)!]^2}\Delta ^{2} C_j.
\end{align}
As shown in Eq. (\ref{Eq31}), $T^{2j+2}$ appears in the expression of QFI together with the variance of the $j$th-order commutator $C_j$ in initial state, which provides a possibility to realize higher-order time scaling in QFI. 

In the following, we will design the Hamiltonian into the form $H_f=fH_0+H_1$ such that an ancilla probe is introduced to the system and does not couple directly with the signal $f$. We will explore the role played by the commutator $[H_0, H_1]$ in generating higher-order term in coherence time in QFI. In particular, we propose two specific models, the qubit-HO Ramsey interferometer and a chain of coupled harmonic oscillator to investigate the sensitivity enhancement.

\begin{figure}
\centering
\includegraphics[width=0.45\textwidth]{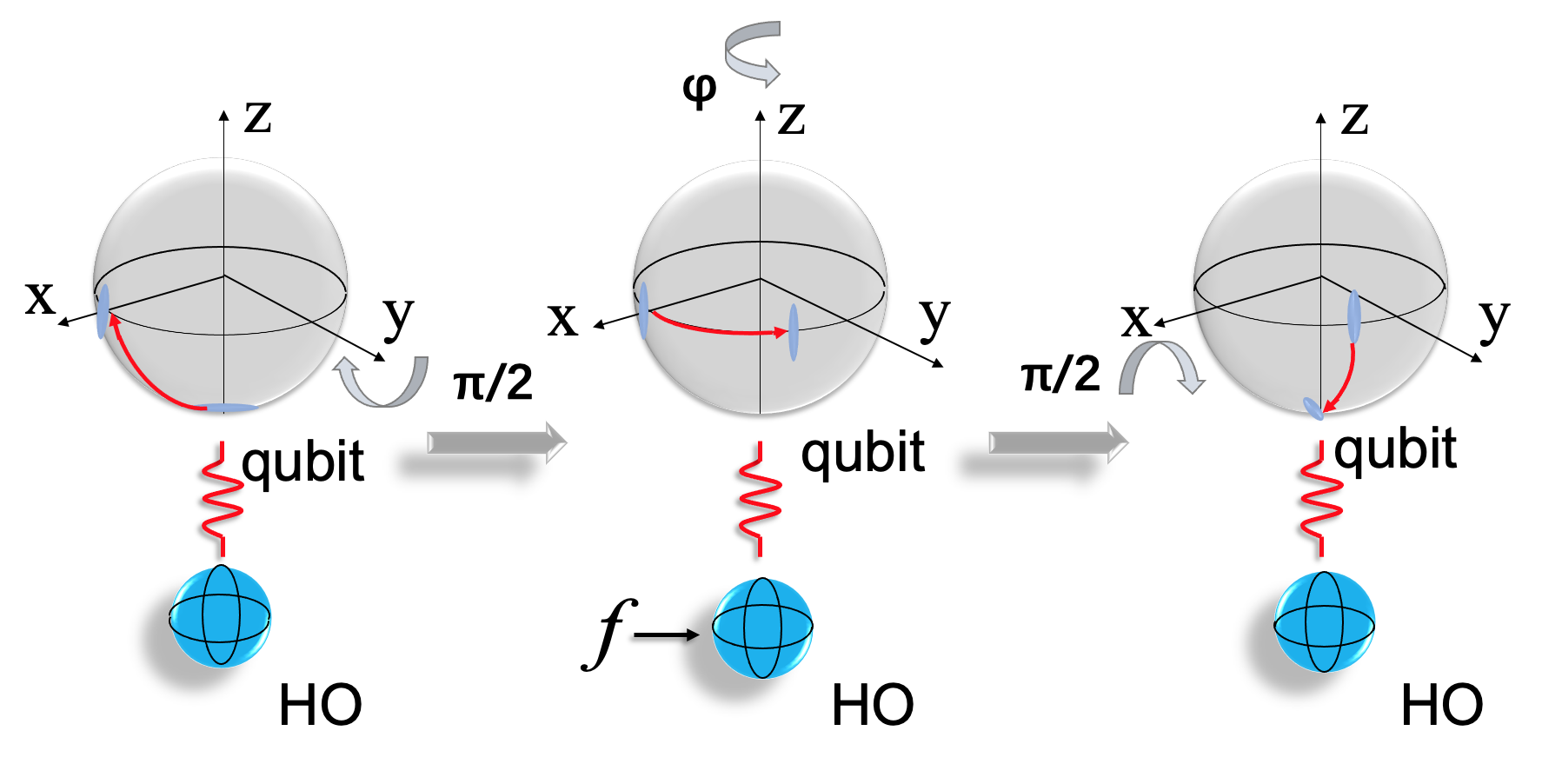}
\caption{The qubit-HO Ramsey interferometer: the system is composed of a qubit coupled with a HO through the force $f$ acting on the HO. The qubit rotates $\pi/2$ along the $x$-axis firstly. Then $f$ is encoded on the HO and the system evolves under $H_f$, which makes the qubit rotate $\phi$ along the $z$-axis. Lastly, the qubit rotates $\pi/2$ along the $y$-axis.}
\label{Fig1}
\end{figure}

\emph{Achieving quartic time scaling in QFI.} --- 
We start our analysis with a qubit-HO Ramsey interferometry model. As shown in Fig. \ref{Fig1}, the initial state of the qubit and the HO is prepared into a separable state $\ket{\varphi_0}=\ket{\psi_0}\otimes \ket{0}$, where $\ket{\psi_0}$ is the initial state of HO and $\ket{0}$ ($\ket{1}$) is the ground (excited) state of the qubit. The qubit is subsequently subjected to a $\pi /2$ pulse along $y$-axis, transforming it into $\frac{1}{{\sqrt 2 }}(\ket{0}  + \ket{1} )$. Afterwards, the system evolves under the Hamiltonian $H_f= fX+g P \sigma _z$ for time $T$, where $X = \frac{1}{\sqrt 2}(a^\dag + a)$, $P = \frac{i}{\sqrt 2}  ({a^\dag } -a)$, $\sigma _z$ is the Pauli-Z gate, $a$ is the annihilation operator and $g$ is the qubit-HO coupling strength.  Finally, another $\pi/2$-pulse around the $x$-axis is applied to the ancilla qubit.	

In order to obtain QFI, we calculate commutators as: $C_0=X$,  $C_1=[(fX+g P \sigma _z), X]=-ig \sigma _z$ and $C_j=0$ for $j \geq 2$. Since the initial state of the system is $\ket{\varphi_0 }=\frac{1}{{\sqrt 2 }}(\ket{0}  + \ket{1} )\ket{\psi_0}$, thus ${\rm Cov}(C_k,C_j)=0$. According to Eq. (\ref{Eq31}) we calculate QFI as:
\begin{align}\label{Eq4}
F_Q=4\Delta ^2(\frac{1}{2}gT^2 \sigma_z+TX)= g^2T^4 \Delta ^2\sigma_z+4T^2 \Delta ^2X
\end{align}
for a separable $\ket{\varphi_0}$. In particular, if $\ket{\psi_0}=\ket{0}$ is the vacuum state, then $F_Q=g^2 T^4+4T^2$; if $\ket{\psi_0}=S(r)\ket{0}$, where $S(r)$ is the squeeze operator and $r>0$, then $F_Q \approx g^2 T^4$ for sufficiently large $r$. In both cases, a $T^{4}$-scaling of QFI can be achieved. With a strong coupling $g$ and relatively long coherent interaction time $T$, a $T^2$ enhancement of $\Delta f$ can be obtained. As shown in Eq.~(\ref{Eq4}), the higher-order term $g^2T^4$ originates from the non-commutativity between $X$ and $P$. Therefore, the non-commutativity in the Hamiltonian could be a useful quantum resource to enhance estimation precision.  

The QCRB only gives the optimal lower bound for $\Delta f$. We still need to show there exits an optimal quantum measurement to saturate this bound~\cite{Zhou2020}. One such optimal measurement strategy is given by the measurement observable $M=\Pi \otimes \sigma_x$ where $\Pi=(-1)^{a^\dag a}$ is the parity operator on the quantum HO, satisfying $\Pi\psi(x)=\psi(-x)$. By preparing the HO in the initial state $\ket{\psi_0}=S(r)\ket{0}$, we can calculate $\Delta f$ via the error propagation formula:
\begin{align*}
    \Delta f = \frac{{\Delta M}}{{| {\frac{{\partial \langle M \rangle}}{{\partial f}}}
    |}}\approx \frac{1}{gT^2}=\frac{1}{\sqrt{F_Q}}, \text{ for sufficiently large }r,
\end{align*}
where $\Delta M$ and $\langle M \rangle$ are the variance and the expectation value of $M$ in the final state $\ket{\varphi_f}$ (See details in Supplementary Materials). Thus, a $T^2$ enhancement of $\Delta f$ can be obtained under such measurement design. 

Furthermore, the non-commutativity and entanglement can be used simultaneously to increase estimation precision. We design a system of $n$ non-interacting HOs, each of which is coupled with the global force $f$, and an ancila qubit, under the Hamiltonian $H_f= f\sum_{k=1}^n X_k+g \sum_{k=1}^n P_k \sigma _z^{(k)}$. By analogy with Eq. (\ref{Eq4}), if the $n$ qubits are in the GHZ state, we can calculate the QFI according to Eq. (\ref{Eq3}):
\begin{align}
 F_Q&=g^2 T^4\Delta ^2(\sum_k^n \sigma_z^{(k)}) + 4T^2\Delta ^2 (\sum_k^n X_k)\nonumber\\
 &=n^2 g^2 T^4 + 4T^2\Delta ^2(\sum_k^n X_k). 
\end{align}
Hence, a quadratic improvement with respect to $n$ is obtained in QFI, multiplied by the quartic time scaling, compared with the scheme where the $n$ qubits are in a separable state. 

\emph{Achieving $T^{2n+2}$ time scaling in QFI.} --- 
Next, we continue the exploration of the choice of the Hamiltonian $H_f$ in order to generate higher-order terms of $T$ in QFI, based on the intuition gained from the qubit-HO Ramsey interferometry model. The first attempt is to choose $H_f=fX^n+g P \sigma _{z}$, which gives: 
\begin{align}
    C_j =\frac{(-ig)^jn!}{(n-j)!} X^{n-j}\sigma_{z}^j.
\end{align}
in Eq.~(\ref{Eq31}). For odd $n$, and for initial state $\ket{\varphi_0}=\frac{1}{\sqrt 2 }\ket{\psi_0}(\ket{0}  + \ket{1} )$, where $\ket{\psi_0}=\ket{0}$ is the vacuum state, QFI is further simplified into:
\begin{align*}
F_Q=&4\Delta^2\big(\sum_{j=0}^n \frac{(gT)^{j+1} \binom{n}{j}}{g(j+1)} X^{n-j}\sigma_z^{j} \big)=\frac{4(gT)^{2n+2}}{g^2(n+1)^2}\\
&+\sum_{j,k=0}^{n-1}\frac{4(gT)^{j+k+2}\tbinom{n}{k}\tbinom{n}{j}}{g^2(j+1)(k+1)}\frac{(2n-j-k-1)!!}{(\sqrt{2})^{2n-j-k}}, 
\end{align*}
where $j+k$ is even in the summation. Thus we can achieve $T^{2n+2}$ time scaling in QFI. For instance, for $n=3$, we have 
\begin{align*}
F_Q=\frac{15}{2}T^2+\frac{51}{4}g^2T^4+\frac{7}{2}g^4T^6+\frac{1}{4}g^6T^8. 
\end{align*}
Nevertheless, it is difficult to experimentally implement such Hamiltonian $X^n$. Alternatively, we can design $H_f$ to be a chain of $n$ coupled HOs with common HO-HO interactions to reach higher-order $T$ scaling in QFI.
\begin{figure}
\centering
\includegraphics[width=0.45\textwidth]{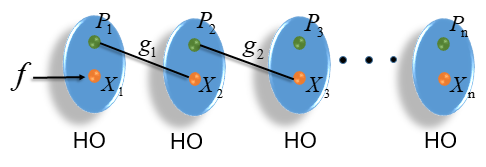}
\caption{A model containing a chain of $n$ coupled HOs, where the signal $f$ couples to the first HO via the interaction $fX_1$, and each HO interacts with nearest neighbors via the interaction $g_{j}P_{j} X_{j+1}$. }
\label{Figue2}
\end{figure}

\begin{figure*}
    \centering
    \subfigure[]{\includegraphics[width=0.4\textwidth]{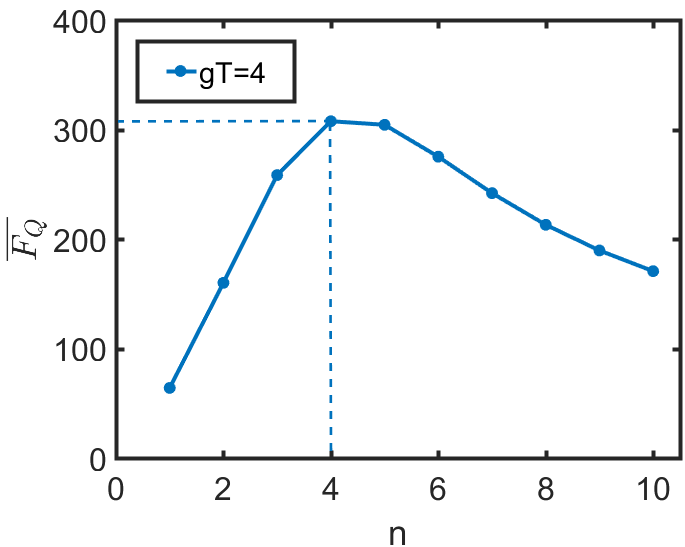}\label{3_1}}
    \subfigure[]{\includegraphics[width=0.4\textwidth]{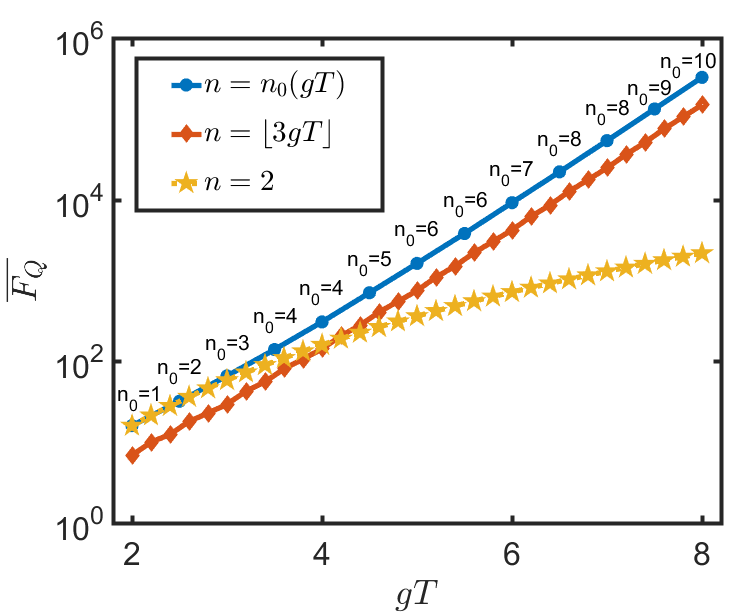}\label{3_2}}
    \caption{Optimal averaged QFI also demonstrates exponential enhancement. In (a), for $gT=4$, the averaged QFI versus $n$ is plotted, with the maximal value achieved at $n=4$. In (b), the curves for $\overline{F_Q}(n_0(gT),gT)$ and $\overline{F_Q}(\lfloor 3 gT \rfloor,gT)$ grow exponentially with $gT$, while the curve for $\overline{F_Q}(n=2,gT)$ grows polynomially with $gT$.    
    }
   \label{Fig3}
\end{figure*}

\emph{Achieving exponential enhancement of QFI.} --- 
As shown in Fig. \ref{Figue2}, we design a chain of $n$ coupled HOs, where the parameter $f$ is coupled with the first HO, and each HO interacts only with its adjacent neighbors, characterized by the total Hamiltonian $H_{f}=fH_0+H_1=fX_1+\sum^{n-1}_{j=1} g_{j}P_{j} X_{j+1}$, with $g_j$ as the coupling constant, and $X_{j}$ and $P_{j}$ as the quadratures of the $j$th HO. The signal $f$ to be estimated is coupled to the first HO through $X_1$, and we introduce $n-1$ additional HOs as an ancila for better measurement precision. This model is also known as the bosonic Kitaev-Majorana chain~\cite{PhysRevX.8.041031}. Analogous to the previous setup, after the evolution $U_f=e^{-i(fH_{0}+H_1)T}$, we can define $h$ and calculate the commutators:
\begin{align}
    C_j=[H_f,C_{j-1}]=(-ig)^{j}X_{j+1}.
\end{align}
which are local operators on different subsystems. Hence, if the entire system is prepared in a separable initial state, we have ${\rm Cov}(C_k,C_j)=0$. In order to analyze the expression of $F_Q$, we further assume all HOs are prepared in the same initial state, which gives $\Delta ^{2} X_{j+1}=\Delta ^{2} X$. In this case Eq. (\ref{Eq31}) is reduced to 
\begin{align}\label{ieqnPQ}
F_Q =\frac{4\Delta ^{2} X}{g^2}\sum\limits_{j=1}^{n}\Big(\frac{(gT)^{j}}{j!}\Big)^2,
\end{align}
One can further show the following inequality for $gT\ge 2$ and $n=\lfloor 3 gT \rfloor$: 
\begin{align}\label{ieqnieqnPQ}
    {F_Q} &=\frac{4\Delta ^{2} X}{g^{2}}\sum\limits_{j=1}^{\lfloor 3 gT \rfloor} \Big(\frac{(gT)^{j}}{j!}\Big)^2   > \frac{2\Delta ^{2} X}{g^{2}} e^{gT}.
\end{align}
 Detailed proof of Eq. (\ref{ieqnieqnPQ}) can be found in the appendix. The intuition behind the proof is, the expression of $F_Q$ in Eq. (\ref{ieqnieqnPQ}) is very similar to the first $n$ terms in the Taylor expansion of $e^{gT}$. The key point is, given the value of $gT$, one needs to determine how large $n$ has to be so that $F_Q$ can scale as $e^{gT}$ up to a constant. It turns out that $n=\lfloor 3 gT \rfloor$ is sufficient to make QFI scale exponentially with respect to $gT$, as long as $gT\ge 2$. Notice that such exponential enhancement of QFI only requires $n-1$ number of HO-HO nearest-neighbor interactions among $n$ HOs, which is crucial to justify the efficiency and the effectiveness of the proposal for exponential enhancement. In addition, we can also design an optimal measurement to saturate the QCRB for such exponential scaling of QFI. Specifically, we can choose the parity operator $M=\otimes_{k=1}^{n} \Pi_k$ as the measurement observable to reach the exponential enhancement of $\Delta f$ when $n$ and $T$ grows linearly under the condition $n=\lfloor 3 gT \rfloor$~\cite{Wang2021}. 
 


\emph{Achieving the optimal averaged QFI for exponential enhancement} --- 
It turns out that $n=\lfloor 3 gT \rfloor$ is only sufficient but not necessary or efficient to obtain exponential enhancement of QFI in the previous model. In fact, we can define the averaged QFI per HO as $\overline{F_Q}(n, gT)=F_Q/n$, which is a function of both $n$ and $gT$. For a fixed $gT$, $\overline{F_Q}$ is found to first increase and then decrease as $n$ grows, as shown in Fig.~\ref{Fig3}(a); in other words, there exists an optimal value of $n=n_0(gT)$ to reach to the maximum of $\overline{F_Q}$ for fixed $gT$. One can then numerically find the value $n_0(gT)$ for different values of $gT$. Hence, $\overline{F_Q}(n_0(gT),gT)$ can be plotted as a curve against $gT$ in Fig.~\ref{Fig3}(b), and in the plot it seems to grow exponentially with $gT$. For comparison, we also plot $\overline{F_Q}(\lfloor 3 gT \rfloor,gT)$ and $\overline{F_Q}(n=2,gT)$ in the same figure. Analogous to the previous discussion, one can rigorously prove that for $gT\ge 2$ and $n=\lfloor 3 gT \rfloor$, $\overline{F_Q}$ grows exponentially with $gT$, as demonstrated in Fig.~\ref{Fig3}(b); nevertheless, the corresponding $\overline{F_Q}$ is far from being optimal, compared to the $\overline{F_Q}$ curve for $n=n_0(gT)$. Hence, the $n=\lfloor 3 gT \rfloor$ condition is only helpful to construct the rigorous proof for exponential enhancement, but not necessary. In practice, $n=n_0(gT)$ should be enough to achieve the exponential behavior. Moreover, keeping $n$ growing with $gT$ is crucial to obtain the exponential enhancement; for a fixed value of $n$, $\overline{F_Q}$ only grows polynomially against $gT$, as illustrated in Fig.~\ref{Fig3}(b). 

 \emph{Conclusion.}---As a type of quantum resource, coherent interaction time plays a crucial role in quantum precise measurement. By introducing an auxiliary system which do not couple directly to the signal to be estimated, we can express QFI as a power series in coherence time. For the qubit-oscillator Ramsey interferometer model, QFI has a quartic time scaling; for a chain of coupled harmonic oscillators, with the number of coupling terms growing linearly with time, the corresponding QFI is shown to have an exponential time scaling. Our results suggest that linear scaling in both time and the number of coupling terms is sufficient to obtain exponential enhancement in continuous-variable quantum metrology. 

This research was supported by the National Key R$\&$D Program of China, Grant No. 2018YFA0306703.

\bibliographystyle{apsrev4-1}
\bibliography{Ref}

\clearpage

\onecolumngrid

\appendix


\label{sec:Intro}
\section*{Measurement design to achieve $1/T^2$ scaling in measurement uncertainty in the qubit-harmonic-oscillator model}

In this work, we use the following definition of the quadratures, 
\begin{align*}
X&\equiv\frac{1}{\sqrt{2}}(a+a^\dag)\\
P&\equiv\frac{i}{\sqrt{2}}(a^\dag-a )
\end{align*}
satisfying $[X,P]=i$. 

Let's consider a quantum harmonic oscillator(HO) system coupled with a signal $f$, under the Hamiltonian $H_0=fX$. Our aim is to measure $f$ as accurately as possible. We introduce a probing qubit to interact with the HO through the Hamiltonian $H_1=g \sigma_z \otimes P$. The total Hamiltonian of this qubit-HO model is $H_f=H_0+H_1=fX+g\sigma_z P$. Let the initial state of the qubit-HO system to be $\ket{\varphi_0}$, its unitary evolution after time $T$ is described by:
\begin{align*}
U_f=e^{-iT(fX+gP\sigma_z)}&=e^{-iTgP\sigma_z}e^{-iTfX}e^{-\frac{i}{2}fgT^2\sigma_z}=e^{-iTgP\sigma_z-\frac{i}{2}fgT^2\sigma_z}e^{-iTfX}
\end{align*}
Thus, the information about $f$ is encoded into $U_f$. If we choose the initial state as a separable state $\ket{\varphi_0}=\frac{1}{\sqrt{2}}(\ket 0+\ket 1)\ket {\psi_0}$, then the final state after $U_f$ becomes:
\begin{align*}
\ket{\varphi_f}=U_f\ket{\varphi_0}&= \frac{1}{\sqrt{2}} e^{-iTgP\sigma_z-\frac{i}{2}fgT^2\sigma_z}(\ket{0}+\ket{1}) e^{-iTfX}\ket {\psi_0}\\
&= \frac{1}{\sqrt{2}}\Big( e^{-iTgP-\frac{i}{2}fgT^2}\ket 0+ e^{iTgP+\frac{i}{2}fgT^2}\ket 1\Big) e^{-iTfX}\ket{\psi_0}\\
&= \frac{e^{-\frac{i}{2}fgT^2}}{\sqrt{2}}\Big( \ket{0} e^{-iTgP}e^{-iTfX}\ket{\psi_0}+ e^{ifgT^2}\ket{1} e^{iTgP} e^{-iTfX}\ket{\psi_0}\Big)
\end{align*}

The corresponding QFI is:
\begin{align*}
    F_Q&=4\Delta^2(\sum_{j=0}^\infty\frac{i^jT^{j+1}}{(j+1)!}C_j)=4\Delta^2(TX+\frac{1}{2}T^2g\sigma_z)=4T^2 \Delta^2 X+g^2T^4 \Delta^2\sigma_z
\end{align*}
where $C_0=X$ and $C_1=-ig\sigma_z$, and $C_j = 0 $ for $j\ge 2$. In particular, if we further choose the initial state of the HO to be $\ket{\psi_0} = \ket{x=0}$, an eigenstate of $X$ at position $x=0$, then $\ket{\varphi_0}=\frac{1}{\sqrt{2}}(\ket 0+\ket 1)\ket{x=0}$, $\Delta^2 X=0$, $\Delta^2\sigma_z=1$, and
\begin{align*}
F_Q&=4T^2 \Delta^2 X+g^2T^4 \Delta^2\sigma_z= g^2T^4
\end{align*}
Hence the minimum measurement uncertainty for such $\ket{\psi_0}$ is $\min\Delta f=\frac{1}{\sqrt{F_Q}}=\frac{1}{gT^2}$. 

Next, we show there does exist a quantum measurement that can achieve such minimum measurement accuracy. Let's choose $M=\sigma_x\otimes \Pi$ where $\Pi$ is the parity operator on the quantum HO satisfying $\Pi\psi(x)=\psi(-x)$. It turns out that $\Pi=e^{i\pi a^\dag a}=(-1)^{a^\dag a}$. 

For the initial state $\ket{\varphi_0}=\frac{1}{\sqrt{2}}(\ket 0+\ket 1)\ket{x=0}$, the final state becomes:
\begin{align*}
\ket{\varphi_f}=U_f\ket{\varphi_0}=\frac{e^{-\frac{i}{2}fgT^2}}{\sqrt{2}}(\ket{0}\ket{gT}+e^{ifgT^2}\ket{1}\ket{-gT})
\end{align*}
Hence, the error propagation formula gives: 
\begin{align*}
\langle M \rangle &= \frac{1}{2}(\bra{0}\bra{gT}+e^{-ifgT^2}\bra{1}\bra{-gT})(\sigma_x\otimes \Pi)(\ket{0}\ket{gT}+e^{ifgT^2}\ket{1}\ket{-gT})\\
&=\frac{1}{2}(\bra{0}\bra{gT}+e^{-ifgT^2}\bra{1}\bra{-gT})(\ket{1}\ket{-gT}+e^{ifgT^2}\ket{0}\ket{gT})\\
&=\frac{1}{2}(e^{-ifgT^2}+e^{ifgT^2})=\cos(fgT^2)\\
\langle M^2 \rangle &=1\\
\Delta^2 M&=\sin(2fgT^2)\\
\Delta_f&=\frac{\Delta M}{|\partial_f  \langle M \rangle|}=\frac{|\sin(fgT^2)|}{gT^2\sin(fgT^2)|}=\frac{1}{gT^2}=\frac{1}{\sqrt{F_Q}}
\end{align*}

In practice, we cannot exactly prepare the HO in the unphysical state $\ket{x=0}$; instead, we can prepare the HO in a squeezed state centered as the zero position to approximate $\ket{x=0}$. In the case, the final state after $U_f$ becomes
\begin{align*}
\ket{\varphi_f}&= \frac{e^{-\frac{i}{2}fgT^2}}{\sqrt{2}}\Big( \ket{0} e^{-igPT}\ket{\alpha_0}+ e^{ifgT^2}\ket{1} e^{igPT} \ket {\alpha_0}\Big)\\
&=\frac{1}{\sqrt{2}}(\ket{0}\ket{gT}+e^{ifgT^2}\ket{1}\ket{-gT})
\end{align*}

For Fock state $\ket{n}$, in the position representation, $\psi_n(x)\equiv n(x)=\langle x \ket{n}$. $\psi_n$ are real functions: $\R\to \R$. $\psi_n$ is an even-function for an even $n$, and an odd-function for an odd $n$. Hence, $\hat\Pi \psi_n(x)=\psi_n(-x)=(-1)^n \psi_n(x)$. 
\begin{align*}
\alpha(x)&=\langle x\ket{\alpha}= e^{-\frac{|\alpha|^2}{2}}\sum_n  \frac{\alpha^n}{\sqrt{n!}} \psi_n(x) \\
\hat\Pi  \alpha(x)&=e^{-\frac{|\alpha|^2}{2}}\sum_n  \frac{\alpha^n}{\sqrt{n!}} \hat\Pi\psi_n(x)=e^{-\frac{|\alpha|^2}{2}}\sum_n  \frac{(-\alpha)^n}{\sqrt{n!}}\psi_n(x)=(-\alpha)(x)
\end{align*}

Hence, 
\begin{align*}
\hat\Pi  \ket{\alpha}&=\ket{-\alpha}
\end{align*}

The Displacement operator is given by
\begin{align*}
    D(\alpha) = e^{\alpha \hat{a}^\dagger - \alpha^\ast \hat{a}}
\end{align*}
A few important facts: 
\begin{align*}
e^{-igTP}&=e^{\frac{1}{\sqrt{2}}gT(a^\dag -a )}=D(\frac{1}{\sqrt{2}}gT)\\
e^{-ifTX}&=e^{-\frac{i}{\sqrt{2}}fT(a^\dag +a )}=D(-\frac{i}{\sqrt{2}}fT)\\
D(\alpha)D(\beta)&=e(\alpha\beta^\ast-\alpha^\ast\beta)D(\alpha+\beta)\\
D(\frac{1}{\sqrt{2}}gT)D(-\frac{i}{\sqrt{2}}fT)&=D(\frac{1}{\sqrt{2}}(gT-ifT))e^{\frac{i}{2}fgT^2}\\
D(-\frac{1}{\sqrt{2}}gT)D(-\frac{i}{\sqrt{2}}fT)&=D(-\frac{1}{\sqrt{2}}(gT+ifT))e^{-\frac{i}{2}fgT^2}
\end{align*}

Hence, the final state: 
\begin{align*}
 \ket{\varphi_f}=&\frac{1}{\sqrt{2}}\Big( \ket{0} e^{-\frac{i}{2}fgT^2}e^{-igPT}e^{-ifTX}\ket{\phi_0}+ \ket{1} e^{\frac{i}{2}fgT^2}e^{igPT} e^{-ifTX}\ket{\phi_0}\Big)\\
 =& \frac{1}{\sqrt{2}}\Big( \ket{0} e^{-\frac{i}{2}fgT^2}D(\frac{1}{\sqrt{2}}gT)D(-\frac{i}{\sqrt{2}}fT)\ket{\phi_0}+ \ket{1} e^{\frac{i}{2}fgT^2}D(-\frac{1}{\sqrt{2}}gT)D(-\frac{i}{\sqrt{2}}fT)\ket{\phi_0}\Big)\\
 =& \frac{1}{\sqrt{2}}\Big( \ket{0} D(\frac{1}{\sqrt{2}}(gT-ifT))\ket{\phi_0}+ \ket{1} D(\frac{1}{\sqrt{2}}(-gT-ifT))\ket{\phi_0}\Big)
\end{align*}

If we choose the initial state of HO to be the squeezed vacuum state $\ket{\phi_0}=S\ket{0}$, then the final state becomes:
\begin{align*}
 \ket{\varphi_f}=\frac{1}{\sqrt{2}}\Big( \ket{0} D(\frac{1}{\sqrt{2}}(gT-ifT))\ket{\phi_0}+ \ket{1} D(\frac{1}{\sqrt{2}}(-gT-ifT))\ket{\phi_0}\Big)
\end{align*}
For $M=\sigma_x\otimes \Pi$, we have:
\begin{align*}
M \ket{\psi_f}&=\frac{1}{\sqrt{2}}\Big( \ket{1} D(\frac{1}{\sqrt{2}}(-gT+ifT))\ket{\phi_0}+ \ket{0} D(\frac{1}{\sqrt{2}}(gT+ifT))\ket{\phi_0}\Big)\\
\bra{\psi_f}&=\frac{1}{\sqrt{2}}\Big( \bra{0} \bra{\phi_0}D(\frac{1}{\sqrt{2}}(-gT+ifT))+ \bra{1} \bra{\phi_0}D(\frac{1}{\sqrt{2}}(gT+ifT))\Big)\\
\langle M \rangle &= \frac{1}{2}\Big(\bra{0} \bra{\phi_0} D(\frac{1}{\sqrt{2}}(-gT+ifT))+ \bra{1} \bra{\phi_0}D(\frac{1}{\sqrt{2}}(gT+ifT))\Big)\\
&\times\Big( \ket{1} D(\frac{1}{\sqrt{2}}(-gT+ifT))\ket{\phi_0}+ \ket{0} D(\frac{1}{\sqrt{2}}(gT+ifT))\ket{\phi_0}\Big)\\
&=\frac{1}{2}(\bra{\phi_0} D(\frac{1}{\sqrt{2}}(-gT+ifT))D(\frac{1}{\sqrt{2}}(gT+ifT))\ket{\phi_0}+\bra{\phi_0}D(\frac{1}{\sqrt{2}}(gT+ifT)))D(\frac{1}{\sqrt{2}}(-gT+ifT))\ket{\phi_0}\\
&=\frac{1}{2}(e^{-ifgT^2}+e^{ifgT^2})\bra{\phi_0}D(\sqrt{2}ifT)\ket{\phi_0}\\
&=\cos(fgT^2)\bra{\phi_0}D(\sqrt{2}ifT)\ket{\phi_0} =\cos(fgT^2)e^{-\sqrt{2}f^2T^2e^{-r}}
\end{align*}
where we have used: 
\begin{align*}
K&\equiv\bra{\phi_0}D(\sqrt{2}ifT)\ket{\phi_0} = \bra{0}S^\dag e^{\sqrt{2}ifT(a^\dag+a)}S\ket{0}=\bra{0} e^{\sqrt{2}ifTS^\dag(a^\dag+a)S}\ket{0}\\
&=\bra{0} e^{\sqrt{2}ifTe^{-r}(a^\dag+a)}\ket{0}=\langle{0}\ket{\alpha=\sqrt{2}ifTe^{-r}}=e^{-\sqrt{2}f^2T^2e^{-r}}\\
\ket{\alpha}&=e^{\frac{|\alpha|^2}{2}}\sum_{n=0}^{+\infty} \frac{\alpha^n}{\sqrt{n!}}\ket{n}\\
K&= \langle 0\ket{\alpha}=e^{-\frac{|\alpha|^2}{2}}=e^{-\sqrt{2}f^2T^2e^{-r}}
\end{align*}

Also we have: 
\begin{align*}
M^2 &=I\\
\langle M^2 \rangle &=1\\
\Delta^2 M&=1-\cos^2(fgT^2)e^{-2f^4T^4e^{-2r}}\\
\frac{\Delta M}{|\partial_f  \langle M \rangle|}&=\frac{\sqrt{1-\cos^2(fgT^2)e^{-2f^4T^4e^{-2r}}}}{|gT^2\sin(fgT^2)K-2\sqrt{2}fT^2e^{-r}e^{-2f^2T^2e^{-r}}\cos(fgT^2)|},
\end{align*}

For a given value $T$, there exists a sufficiently large $r$ such that $e^{-2f^4T^4e^{-2r}}\approx1$, $e^{-r}\approx 0$, $K\approx 0$ and 
\begin{align*}
\Delta f&= \frac{\Delta M}{|\partial_f  \langle M \rangle|}=\frac{\sqrt{1-\cos^2(fgT^2)e^{-2f^4T^4e^{-2r}}}}{|gT^2\sin(fgT^2)K-2\sqrt{2}fT^2e^{-r}e^{-2f^2T^2e^{-r}}\cos(fgT^2)|}\approx \frac{|\sin(fgT^2)|}{gT^2|\sin(fgT^2)|}=\frac{1}{gT^2}=\frac{1}{\sqrt{F_Q}}.
\end{align*}

Thus, we have achieved the $T^2$ enhancement in $\Delta f$.

\section*{Proof of the exponential enhancement}

Define the following quantities: 
\begin{align*}
P(x,n) &\equiv \sum_{j=0}^n \frac{x^j}{j!}\\
Q(x,n)&\equiv \sum_{j=0}^n (\frac{x^j}{j!})^{2}
\end{align*}

We hope to find an appropriate dependence of $n$ on $x$ such that $Q(x,n)$ scales like $e^x$ as $x$ increases. First of all, $P(x,n)$ is the Taylor expansion of $e^{x}$, satisfying $P(x,n)\to e^{x}$ as $n\to +\infty$, for any $x$. Moreover, the larger $x$, the larger $n$ is needed for $P(x,n)$ to approximate $e^x$. One interesting question is, how large $n$ should be so that $P(x,n)$ can can approximate $e^x$ reasonably well, i.e., the difference between $P(x,n)$ and $e^x$ can be made arbitrarily small? 

By Stirling's formula and Cauchy–Schwarz inequality, it is easy to show that, for $x\ge 3$, $n=x^2$ is sufficient to make $|P(x,n=x^2)-e^x|$ sufficiently small and $Q(x,n=x^2)> e^x$. The next question is whether we can further reduce such quadratic dependence of $n$ on $x$ to a linear dependence. The answer is yes, as we have the following lemma:

\begin{lemma} \label{lemma1}
For $x\ge 2$, if $n=3x$ is an integer, we have

(1) $e^x - P(x,n)<\frac{x}{\sqrt{2\pi}}\big(\frac{e}{3}\big)^{3x+1}$;

(2) $Q(x,n)> e^x$.

 \end{lemma}
 
\begin{proof}

For $x>1$, and $n>x$, we have
\begin{align*}
e^x - P(x,n) &= \sum_{j=n+1}^{+\infty} \frac{x^j}{j!}\\
&= \frac{x^{n+1}}{(n+1)!} \left( 1+ \frac{x}{n+2} + \sum_{j=2}^{+\infty}\frac{(n+1)!x^j}{(n+j+1)!} \right)\\
&\leq\frac{x^{n+1}}{(n+1)!} \left( 1+ \frac{x}{n+2} +  \sum_{j=2}^{+\infty}\frac{x^2}{(n+j)(n+j+1)} \right)\\
&=\frac{x^{n+1}}{(n+1)!} \left( 1+ \frac{x}{n+2} + \frac{x^2}{n+2}\right)
\end{align*}

Then, if we choose $n=3x$, and for $x\ge \frac{\sqrt{5}+1}{2}$, we have
\begin{align*}\label{ieqnex}
e^x - P(x,n)  \leq \frac{x^{3x+1}}{(3x+1)!}\left(1+\frac{x}{3x+2}+\frac{x^2}{3x+2}\right)\leq \frac{x^{3x+1}}{(3x+1)!} x
\end{align*}
According to Stirling’s Formula
\[
n!\ge \sqrt{2\pi} n^{n+\frac{1}{2}}e^{-n},
\]
we have
\[
(3x+1)! \ge \sqrt{2\pi} (3x+1)^{(3x+1+\frac{1}{2})}e^{-(3x+1)}.
\]
Thus, we obtain
\[
e^x - P(x,n=3x)  \leq  \frac{x^{3x+1}e^{3x+1}}{ \sqrt{2\pi}(3x+1)^{(3x+1+\frac{1}{2})}} x < \frac{1}{ \sqrt{2\pi}}\left(\frac{ex}{3x+1}\right)^{3x+1} x<  \frac{x}{ \sqrt{2\pi}} \left(\frac{e}{3}\right)^{3x+1},
\]
implying that for sufficiently large $x$, the difference between $P(x,n=3x)$ and $e^{x}$ can be made arbitrarily small. Furthermore, it is easy to show that, for $x\ge 2$,
\begin{align*}
P(x,n=3x)  \ge  e^x -\frac{x}{ \sqrt{2\pi}}\left(\frac{e}{3}\right)^{3x+1}> \frac{10}{11} e^x.
\end{align*}

From Cauchy–Schwarz inequality, we have:
\begin{align*}\label{ieqnPQ}
P(x,n)^2=\Big(\sum_j^{n} \frac{x^j}{j!} \Big)^2 \le n\sum_j^n \Big(\frac{x^j}{j!}\Big)^2=nQ(x,n).
\end{align*}
which gives 
\begin{align*}
P(x,n=3x)^2 &\leq 3x Q(x,n=3x)\\
Q(x,n=3x) &\geq \frac{P(x,n=3x)^2}{3x} > \frac{1}{3x}\left(\frac{10e^x}{11}\right)^2. 
\end{align*}
On the other hand, for $x\ge 2$, 
\begin{align*}
 \frac{1}{3x}\left(\frac{100e^x}{121}\right)&> 1\\
 Q(x,n=3x) &\ge \frac{P(x,n=3x)^2}{3x} >  \frac{1}{3x}\left(\frac{100e^x}{121}\right)e^x > e^x,
\end{align*}
which completes the proof. 

\end{proof}

Thus, we have shown that, for $n=3x$ to be an integer and $x\ge 2$,
\begin{align*}
 Q(x,n=3x) = \sum_{j=0}^{3x} (\frac{x^j}{j!})^{2} > e^x,
\end{align*}

When $n=3x$ is not an integer, we have: 

\begin{align*}
 Q(x,n= \lfloor 3x\rfloor) = \sum_{j=0}^{\lfloor 3x\rfloor} (\frac{x^j}{j!})^{2}\ge \sum_{j=0}^{\lfloor 3x\rfloor} (\frac{(\frac{\lfloor 3x\rfloor}{3})^j}{j!})^{2} > e^{\frac{\lfloor 3x\rfloor}{3}}\ge e^{x-\frac{1}{3}},
\end{align*}
Hence, no matter $3x$ is an integer or not, we have:
\begin{align*}
 Q(x,n=\lfloor 3x\rfloor) = \sum_{j=0}^{\lfloor 3x\rfloor} (\frac{x^j}{j!})^{2} > e^{x-\frac{1}{3}}.
\end{align*}

\begin{theorem}

For $x\ge 2$,
\begin{align*}
\sum_{j=1}^{\lfloor 3x\rfloor} (\frac{x^j}{j!})^{2} >\frac{1}{2}e^x
\end{align*}

\end{theorem}
\begin{proof}
For $x\ge 2$, since $e^{x-\frac{1}{3}}-1> \frac{1}{2}e^x$, we have
\begin{align*}
\sum_{j=1}^{\lfloor 3x\rfloor} (\frac{x^j}{j!})^{2} =Q(x,n=\lfloor 3x\rfloor)-1 > e^{x-\frac{1}{3}}-1>\frac{1}{2}e^x
\end{align*}
\end{proof}

From Eq. (11), we have
\begin{align*}
    {F_Q} &=4\sum_{j=0}^{n-1} \frac{(gT)^{2j+2}}{g^{2}[(j+1)!]^{2}} (\Delta ^{2} X_{j+1})=\frac{4\Delta ^{2} X}{g^{2}}\sum_{j=1}^{n} \Big(\frac{(gT)^{j}}{j!}\Big)^2
\end{align*}   
Substituting $x=gT$ into the expression of $Q(x,n=\lfloor 3x\rfloor)$ and for $gT\ge 2$, we find:
\begin{align*}
Q(x=gT,n=\lfloor 3x\rfloor)&=\sum_{j=1}^{\lfloor 3 gT \rfloor} \Big(\frac{(gT)^{j}}{j!}\Big)^2   > \frac{1}{2} e^{gT}
\end{align*}
Hence, for $gT\ge 2$ and $n=\lfloor 3 gT \rfloor$,  we have 
\begin{align*}
 {F_Q} &=\frac{4\Delta ^{2} X}{g^{2}}\sum\limits_{j=1}^{\lfloor 3 gT \rfloor} \Big(\frac{(gT)^{j}}{j!}\Big)^2   > \frac{2\Delta ^{2} X}{g^{2}} e^{gT}
\end{align*}

Moreover, for $n=3x$ to be an integer and $x\ge 2$, analogous to the above discussion, we can prove:
\begin{align*}
 \frac{Q(x,n=3x)}{n} = \frac{1}{3x}\sum_{j=0}^{n=3x} (\frac{x^j}{j!})^{2} > \frac{e^x}{3x}>\frac{e^x}{6},
\end{align*}
For the averaged QFI, $\overline{F_Q}(n, gT)\equiv \frac{1}{n}F_Q$, we have:
\begin{align*}
\overline{F_Q}(n=\lfloor 3 gT \rfloor, gT)\equiv \frac{1}{n}F_Q(n=\lfloor 3 gT \rfloor, gT) = \frac{4\Delta ^{2} X}{g^{2}}\sum\limits_{j=1}^{\lfloor 3 gT \rfloor} \frac{1}{\lfloor 3 gT \rfloor}\Big(\frac{(gT)^{j}}{j!}\Big)^2 >\frac{(\Delta ^{2} X) e^{gT}}{3g^{2}},
\end{align*}
which also grows exponentially with $gT$.

\end{document}